\documentclass[11pt]{article}
\usepackage[T1]{fontenc}
\usepackage{lmodern}
\usepackage{amsmath,amssymb}
\usepackage{booktabs,longtable,array,calc}
\usepackage{graphicx}
\usepackage{geometry}
\usepackage{microtype}
\usepackage{xcolor}
\usepackage{hyperref}
\usepackage{xurl}
\usepackage{seqsplit}
\definecolor{rimsnavy}{HTML}{173A5E}
\definecolor{rimsslate}{HTML}{526171}
\hypersetup{colorlinks=true,linkcolor=rimsnavy,citecolor=rimsnavy,urlcolor=rimsnavy,pdfborder={0 0 0},pdftitle={More Context, Same Budget: Dual-Bounded Relational Recall Beyond Top-K Retrieval},pdfauthor={Thomson D. Nguy}}
\newcommand{\rimsrule}{\par\noindent\textcolor{rimsslate}{\rule{\linewidth}{0.5pt}}\par}
\title{\vspace{-2.2em}\textbf{\Large More Context, Same Budget: Dual-Bounded Relational Recall Beyond Top-K Retrieval}}
\author{Thomson D. Nguy\\[0.25em]\small Radiant Institute for Manifold Studies (RIMS)}
\date{}
\begin{document}
\maketitle
\vspace{-1.3em}
\rimsrule
\begin{abstract}
More context does not require a larger retrieval budget. Under the same ceiling, a retrieval system can recover more of the evidence a question requires by following relationships between evidence that flat top-k ranking leaves behind. We test that proposition with Dual-Bounded Relational Recall (DBRR), which allocates a fixed retrieval budget between relevance-selected seeds and bounded graph-adjacent context, against matched flat top-k retrieval using the same relevance-ranking stage and the same maximum number of retrieval units and tokens. The outcome is complete recovery of the official HotpotQA supporting-evidence set for each question. Across 7,405 FullWiki questions, the Primary DBRR allocation increased complete supporting-evidence recovery by 23.8 percentage points over its matched flat baseline (paired risk difference 0.2377; question-level bootstrap 95\% interval 0.2269 to 0.2489). It improved 1,952 questions, tied on 5,261, and harmed 192. Bridge questions drove the effect, with a 28.7-point increase; comparison questions showed a smaller 4.2-point difference. In a prespecified, evaluation-only diagnostic population, real relationships also outperformed random-neighbor and degree-preserving shuffled-graph controls. The result is straightforward: under the same context budget, complete-evidence retrieval depends not only on which items rank highest, but on how context is allocated around them. Relational allocation recovered complete evidence sets that flat top-k retrieval left incomplete.
\end{abstract}
\rimsrule
\vspace{0.8em}
\section{Introduction}\label{introduction}

\subsection{Retrieval has an allocation problem}\label{retrieval-has-an-allocation-problem}

A context window is a budget. Once its ceiling is fixed, every item admitted displaces another item that might have been admitted instead. This fact is easy to overlook because retrieval is usually described as ranking: score the candidate material, take the highest-ranked items, and pass them onward. But a ranked list does not spend itself. A system still needs a rule for turning that list into a bounded context.

The simplest rule is to keep taking the next item in rank order. That can work well when the required evidence is independently prominent with respect to the question. It can fail when one piece of evidence is highly relevant to the question but another becomes legible only through its relationship to the first. Under the same ceiling, a system could instead reserve part of the allowance for material adjacent to something it has already found relevant. The first choice spends the budget down the ranking. The second spends part of the budget around selected evidence.

That difference is the subject of this study. It is not a contest between having a graph and having no graph in the abstract. Nor is it a claim that one retrieval architecture should replace all others. It is a narrower experimental question: when the object and token ceilings are matched, can bounded graph-adjacent allocation recover a complete support set more often than spending the same allowance on additional flat-ranked items?

Here, an ``object'' is the counted retrieval unit admitted to context. Matching the object and token ceilings means that both arms faced the same maximum allowance; it does not imply identical realized use or equal end-to-end cost.

HotpotQA provides a useful setting for asking that question because it records supporting facts separately from the answer and distinguishes bridge from comparison questions. In a bridge question, later evidence may become salient through an entity or relation exposed by earlier evidence. In a comparison question, the evidence often supports parallel facts about two subjects. These are benchmark constructions rather than universal cognitive categories, but they make different demands on retrieval visible (Yang et al., 2018; HotpotQA, 2019).

\textbf{Benchmark orientation: not a Study A result.} Consider a published HotpotQA question asking for the former band of the Mother Love Bone member who died shortly before the planned release of \emph{Apple}. One passage identifies Andrew Wood as that member; another identifies Wood as the former lead singer of Malfunkshun. Neither passage alone completes the chain. The retrieval problem is to make both available together (Yang et al., 2018, Fig. 1, p.~2369).

Figure 1 holds the relevance-ranked candidate list and context ceiling fixed while the allocation rule changes. The matched flat baseline continues down the ranking. The Primary relational allocation selects relevant seeds and admits bounded graph-adjacent context. Both routes terminate at the same allowed context ceiling. The figure stops at the cleared public method resolution.

\subsection{The study question}\label{the-study-question}

We evaluated these allocations on the frozen HotpotQA FullWiki development population. The primary outcome was complete supporting-evidence recovery: a binary indicator that all official support required for a question had been recovered. This is intentionally demanding. Recovering one useful sentence is not enough when another official support item remains absent. At the same time, it is still a retrieval outcome. A reader can possess the complete official support and answer incorrectly, or answer correctly without reproducing this retrieval outcome. Study A measures evidence availability, not what a downstream model does with the evidence.

The matched design isolates an allocation contrast after a common relevance- ranking stage. It does not equalize every property of a production system. Graph construction, training, latency, energy, and operational complexity are outside the matched budget and outside the result. The question is therefore neither ``Are graphs cheaper?'' nor ``Does graph retrieval answer questions better?'' It is whether one way of spending the frozen retrieval allowance changes complete support recovery relative to a matched flat control.

\subsection{The result at full strength}\label{the-result-at-full-strength}

It did. Across 7,405 questions, the Primary relational allocation increased complete supporting-evidence recovery by 23.8 percentage points over its matched flat baseline. The paired risk difference was 0.2377, with a 95\% paired question-level bootstrap interval from 0.2269 to 0.2489. Put plainly, the observed difference corresponds to roughly 24 additional complete recoveries per 100 questions under this benchmark design.

The average is not the whole result. The difference was 28.7 percentage points among 5,918 bridge questions and 4.2 points among 1,487 comparison questions. The latter estimate was positive but did not clear the adjusted practical- margin rule and is therefore classified as inconclusive. The Primary allocation also harmed 192 questions that the matched baseline recovered completely. Those facts do not erase the aggregate effect. They tell us what kind of effect it is: large, concentrated, and not monotonic across individual questions.

\subsection{Contributions}\label{contributions}

This paper makes three contributions. First, it provides a paired, matched- budget measurement of complete supporting-evidence recovery under a frozen FullWiki design. The operational ingredients are deliberately simple; the contribution is the isolation and magnitude of the allocation contrast, not a claim to have invented graph-aware retrieval.

Second, it reports the boundary evidence with the headline rather than after it. Prespecified bridge and comparison strata show where the average concentrates. Random-neighbor and shuffled-graph controls test relationship specificity in a restricted diagnostic population. Neutral-labeled sensitivities, improve/tie/harm counts, and an adverse support-resolution anomaly preserve the friction in the result.

Third, it provides independent recomputation with an explicit public-replication boundary. An independent verifier recomputed the reported result and inference from the sealed evidence package. The public method description, however, deliberately omits protected allocation details and does not provide the corpus or reconstruction-capable fixtures required for a source-to-result rerun. Those are different reproducibility claims and are treated separately.

\subsection{Scope}\label{scope}

The result applies to complete support recovery in this frozen experiment. It does not establish downstream answer accuracy, model learning, belief revision, or open-world discovery. The matched flat baseline is an experimental control, not a stand-in for learned path retrieval, multi-hop dense retrieval, GraphRAG, agentic search, web retrieval, or a modern production RAG stack. Within those limits, the result remains consequential: how a finite context budget is spent can materially change whether the complete evidence set is available at all.

\section{Related Work and Field Position}\label{related-work-and-field-position}

\subsection{Lexical relevance ranking}\label{lexical-relevance-ranking}

The common ranking stage in Study A belongs to the BM25 family of lexical ranking methods. BM25 is better understood as a family within the probabilistic relevance tradition than as a single universal implementation. Parameterization, tokenization, document representation, and indexing choices can differ across systems. In this study, the relevant fact is not that a label called ``BM25'' was present; it is that both compared allocations began from the same frozen relevance-ranking stage. The experiment changes what happens after ranking, not the underlying candidate score for one arm alone (Robertson and Zaragoza, 2009).

This distinction separates candidate relevance from context composition. A ranker estimates which items deserve attention individually. A composition rule decides which combination of items will occupy the finite downstream context. The highest-scoring combination is not necessarily the combination that closes an evidence set, particularly when the relevance of a later item is conditional on an earlier item. Study A does not propose a general theory of setwise retrieval, but it treats this post-ranking decision as an experimental variable rather than allowing it to remain implicit.

\subsection{Learned retrieval over Wikipedia graphs}\label{learned-retrieval-over-wikipedia-graphs}

Graph-aware retrieval predates this study. Learned path-retrieval systems have used Wikipedia hyperlinks to search sequential reasoning paths, including cases where later evidence has weak direct lexical relation to the original question. Asai and colleagues, for example, learn recurrent path selection and couple the retriever to a reader. That work demonstrates a richer learned architecture and evaluates a different combination of retrieval and answer-facing outcomes (Asai et al., 2020).

Study A does not reproduce or compete directly with that system. Its relational allocation is frozen rather than learned for the reported comparison; its budgets, graph construction, and endpoint differ; and it stops at complete support recovery. Prior graph-aware retrieval establishes that the primitive is not unprecedented. It does not make the present matched allocation result redundant, nor does the present result establish superiority over learned path retrieval.

\subsection{Multi-hop dense retrieval}\label{multi-hop-dense-retrieval}

Explicit graph traversal is not the only way to retrieve later-hop evidence. Multi-hop dense retrieval can recursively condition a new query representation on passages recovered at an earlier step. This provides a learned route to relationally distributed evidence without following a fixed hyperlink edge. Published MDR work also distinguishes retrieval measures from downstream answer measures, a distinction directly relevant here (Xiong et al., 2021).

Study A did not test MDR. The absence of that comparison is scientifically important because a strong result against a matched flat control does not imply state-of-the-art performance against learned dense retrieval. The present study asks a controlled allocation question; a modern comparator study is a separate experiment.

Learned dense and graph-explicit approaches may also converge behaviorally even when their internal representations differ. A dense retriever conditioned on a first passage can retrieve material related to that passage without traversing an explicit edge. Conversely, an explicit graph rule can reach a related item without learning a new query representation. The present experiment identifies the value of its frozen allocation contrast; it does not establish that explicit edges are uniquely capable of producing the observed advantage.

\subsection{Retrieval-augmented generation and GraphRAG}\label{retrieval-augmented-generation-and-graphrag}

Retrieval-augmented generation couples a retriever to a generator and evaluates what the combined system produces. Evidence recovery may influence that system, but it is not interchangeable with generated-answer accuracy. A generator may ignore available evidence, misuse it, or answer correctly for reasons not captured by complete support recovery. Study A contains no generator outcome (Lewis et al., 2020).

GraphRAG is likewise not a generic name for any retrieval process that touches a graph. A prominent GraphRAG design constructs an LLM-derived entity graph, organizes it into communities, produces summaries, and uses a map-reduce answer workflow for global corpus sensemaking. Study A instead uses a frozen relationship graph for bounded support retrieval in HotpotQA and does not generate answers. The two systems share a broad interest in structure but are different research objects (Edge et al., 2024).

\subsection{Position of Study A}\label{position-of-study-a}

The tested primitive, moving through relationships from relevant material, is not new by itself. The contribution is a strong matched empirical contrast: after a common ranking stage and under matched maximum object and token ceilings, a bounded relational allocation recovered the complete official support set much more often than additional flat-ranked context. The simple control makes the allocation decision visible. The narrow comparator set prevents the result from ranking the method against the modern retrieval field.

This position avoids two unhelpful stories. ``Structure beats ranking'' is too broad for the evidence. ``This is merely links after BM25'' is too dismissive of a large, prespecified, independently verified matched contrast. The scientific claim lies between them: under this frozen design, allocation around relevant seeds was consequential, and the effect's distribution offers testable clues about when it may matter.

The closest unit of contribution is therefore not a new field label but an identified design variable. Existing literatures variously emphasize scoring, iterative query formation, path learning, graph construction, community summarization, or answer generation. Study A makes the allocation that follows initial relevance selection independently visible and measures it with a strict support-set endpoint. That is a modest claim about novelty and a strong claim about the observed effect; the two should not be confused.

\section{Study Design}\label{study-design}

\subsection{Frozen benchmark population}\label{frozen-benchmark-population}

The evaluation population comprised 7,405 questions from the frozen HotpotQA FullWiki development design. In the FullWiki setting, relevant evidence must be located within a resident Wikipedia-derived corpus rather than supplied as gold paragraphs. HotpotQA separately represents official supporting facts and answer targets and provides official bridge and comparison question types. The frozen population contained 5,918 bridge questions and 1,487 comparison questions (Yang et al., 2018; HotpotQA, 2019).

The study used the official data under its stated license and acquisition route. Public materials cite those instructions but do not redistribute dataset, corpus, index, or per-query retrieval bytes. The result is a closed- corpus benchmark result. ``FullWiki'' describes the benchmark retrieval setting; it does not make the experiment a test of open-world discovery.

The question was the unit of analysis. All arm outcomes were joined by the same frozen question identity before paired contrasts were computed. This pairing is central to the design: a difficult question is compared with itself under two allocations rather than with a different question assigned to another arm. Consequently, the primary estimate depends on discordant outcomes: questions for which one allocation recovered the complete support set and the other did not. Questions on which both succeeded and questions on which both failed are ties.

The bridge and comparison counts exhaust the reported full population. They are official benchmark strata, not post hoc semantic annotations supplied by the authors. Other prespecified regime surfaces were retained as secondary heterogeneity checks. None changes the identity of the full-population headline or authorizes removal of questions that were difficult, anomalous, or harmful.

\subsection{Outcome: complete supporting-evidence recovery}\label{outcome-complete-supporting-evidence-recovery}

For each question and retrieval arm, complete supporting-evidence recovery was scored as one only when the retrieved context contained the entire official support set; otherwise it was scored as zero. The primary endpoint is therefore binary and question-paired. It differs from partial supporting-fact recall or F1, where recovering some but not all support can receive credit. It also differs from answer exact match, answer F1, joint metrics, citation quality, and generated-answer correctness (Yang et al., 2018).

This distinction is not semantic housekeeping. Complete support recovery tests whether the full annotated evidentiary substrate was made available under the retrieval ceiling. It does not test whether a downstream reader recognized the right inference, whether the annotations exhaust all sufficient evidence, or whether the system changed a belief.

The endpoint also makes partial and complete failure deliberately coarse. A retrieval containing all but one official support item receives the same binary score as a retrieval containing none of them. That strictness serves the study's question, whether the complete annotated set crossed the context boundary, but it does not measure how close an incomplete retrieval came, how redundant the support was, or how severe a harm case would be for an actual reader. Those distinctions require additional recall, ranking, and downstream outcome surfaces.

\subsection{Common ranking and matched allocations}\label{common-ranking-and-matched-allocations}

All primary comparisons began from a common frozen relevance-ranking stage. The matched flat baseline used its remaining allowance for additional items in that ranking. The relational arm used relevance-selected seeds plus bounded graph-adjacent context. The compared arms were matched on object and token ceilings under the frozen protocol. Thus, the experimental contrast is how the allowed context was allocated, not whether one arm was permitted an arbitrarily larger context window (Robertson and Zaragoza, 2009).

``Matched'' has a precise and limited meaning here. Each relational allocation is compared with its own flat baseline under the corresponding frozen ceilings. It does not mean that all retrieved items had identical length, that both arms consumed identical realized context on every question, performed identical graph-related work, or had equal end-to-end computational cost. Matching removes the simplest maximum-allowance explanation for the retrieval contrast while leaving realized use, system construction, and runtime economics for future measurement.

The public paper uses neutral labels only. ``Primary'' identifies the headline relational allocation and its matched baseline. ``Sensitivity A'' and ``Sensitivity B'' identify two additional relational allocations and their own matched baselines. Seed-only, random-neighbor, shuffled-graph, and evaluation- only oracle surfaces are used only for bounded diagnostics. The neutral labels do not encode a public configuration order.

The public description stops at the cleared high-level allocation contrast. The resulting reproducibility boundary appears later in the manuscript.

\begin{figure}
\centering
\includegraphics[width=0.96\linewidth,height=\textheight,keepaspectratio,alt={Same ceiling, different allocation. A common relevance-ranked candidate list feeds two routes under matched maximum object and token ceilings. The flat route continues down the ranking. The relational route reserves part of the allowance for bounded context adjacent to selected evidence. Both terminate at the same allowed context ceiling; held allocation mechanics are not shown.}]{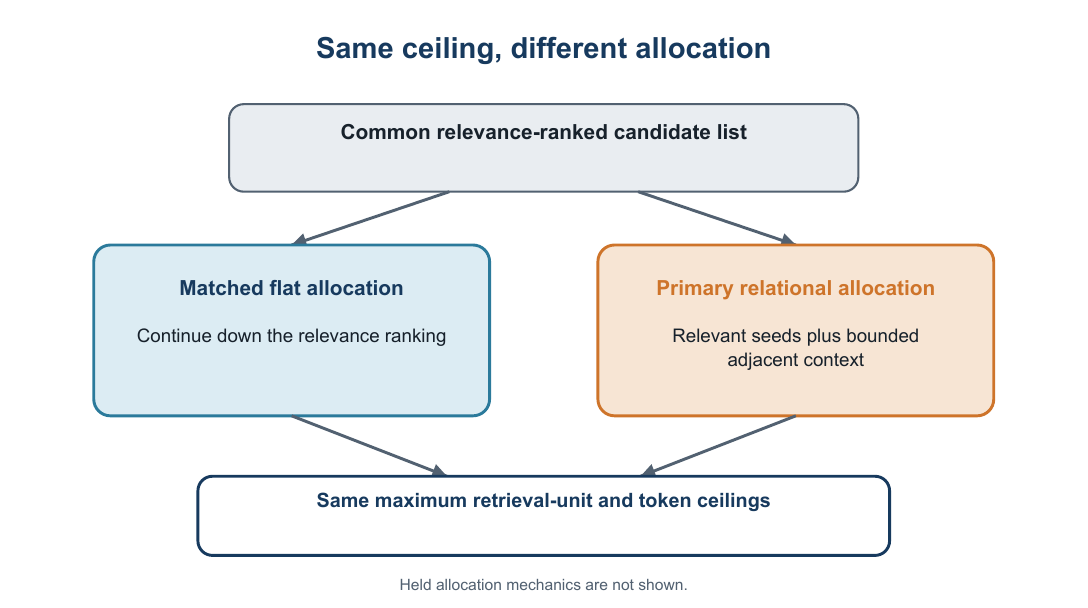}
\caption{Same ceiling, different allocation. A common relevance-ranked candidate list feeds two routes under matched maximum object and token ceilings. The flat route continues down the ranking. The relational route reserves part of the allowance for bounded context adjacent to selected evidence. Both terminate at the same allowed context ceiling; held allocation mechanics are not shown.}\label{fig:figure1}
\end{figure}

\subsection{Frozen graph and provenance}\label{frozen-graph-and-provenance}

The frozen relationship graph contained 5,233,329 nodes and 19,300,783 edges; 3,945,864 links were recorded as dangling under the frozen construction. Its sanitized identity SHA-256 was {\footnotesize\texttt{\seqsplit{728346816e8a84d77af2711802f0a98b468feb83a70cf6f1d21d9504c5b8e201}}}. The sanitized result and retrieval identities were, respectively, {\footnotesize\texttt{\seqsplit{2273f5fc4e6ba73306cfd234010978a25b2944def07050dd0402907027346ce6}}} and {\footnotesize\texttt{\seqsplit{9ba944198a715627b0cb57ccf20aa4f07616ca283717f41752efa8e479e84370}}}.

These values bind the reported analysis to a fixed result, retrieval surface, and graph identity. They do not establish that the graph was optimal, cheap to construct, or representative of a production knowledge graph. No custody locator or reconstruction-capable manifest is included.

At the cleared public resolution, the graph is a fixed relationship structure over retrieval objects in the resident corpus, and ``graph-adjacent'' means that a candidate object is admitted through a recorded relationship to a selected seed. The cleared public description does not resolve which relationship semantics carried the observed effect or would transfer beyond the frozen corpus.

\subsection{Prespecified populations, strata, and diagnostics}\label{prespecified-populations-strata-and-diagnostics}

The full 7,405-question population supplied the Primary headline. Official bridge and comparison question types supplied prespecified strata. These strata were not created after observing the aggregate estimate and must be interpreted together.

A separate opportunity population contained 3,562 questions. Its membership used official support and reachability in the real graph and was defined for evaluation after retrieval outputs were immutable. In ordinary terms, this after-the-fact diagnostic knows where the real graph contained a usable route to gold support. It is therefore gold-derived, conditioned on the real graph, and evaluation-only. It cannot identify suitable questions in deployment. The population supports a conditional comparison among the real relationship graph, a random-neighbor control, and a degree-preserving shuffled-graph control. It does not replace the full-population headline or provide an unconditional causal estimate.

The controls ask different questions. The random-neighbor surface asks whether spending context on neighbor-shaped material is sufficient when the actual relationship is discarded. The degree-preserving shuffled surface retains a coarse structural property while breaking the frozen semantic arrangement. Neither control is a competitive retriever. Their purpose is diagnostic: if the real graph does not separate from them when a known opportunity is present, the relationship-specific account is weakened.

The plan also retained two non-empty, prespecified contained-regime interaction surfaces and one empty hub-oriented regime. These are effect-modification tests, not alternate headline populations. They assess whether the relational-minus- flat contrast changes across a frozen pre-query regime indicator. Because an interaction is a difference between differences, it must not be read as the recovery effect inside either regime by itself.

\subsection{Statistical analysis}\label{statistical-analysis}

For each paired comparison, we estimated the risk difference in complete support recovery: the relational arm's mean binary outcome minus that of its matched flat baseline on the same questions. Paired outcomes were also reported as improve, tie, and harm counts. An improvement is a question recovered completely by the relational allocation but not the baseline; harm is the reverse. The risk difference equals the net discordance divided by the paired population.

Let \(Y_{Ri}\) and \(Y_{Fi}\) denote binary complete-recovery outcomes for the relational and matched flat allocations on question \(i\). The paired risk difference is

\[
\widehat{RD}=\frac{1}{n}\sum_{i=1}^{n}(Y_{Ri}-Y_{Fi})
=\frac{N_{\mathrm{improve}}-N_{\mathrm{harm}}}{n}.
\]

This identity makes the harm accounting part of the estimand rather than a separate anecdotal analysis. Ties contribute zero to the difference, whether both arms succeeded or both failed. Because discordance counts alone do not reveal those two tie components, absolute arm rates were separately summed from the sealed per-question outcomes and recorded in a derivation receipt.

The frozen primary route combined several views of the paired binary contrast. It used an exact two-sided McNemar test on discordant pairs (McNemar, 1947). The headline 95\% interval was a 10,000-resample percentile bootstrap over the paired question-level differences with seed 4401 (Efron, 1979). The stored paired- difference and rank-specific familywise percentile interval surfaces used a separate 10,000-resample procedure with seed 4488. A 90\% interval from two one- sided equivalence tests was compared with the frozen equivalence region {[}-0.03,+0.03{]} (Schuirmann, 1987).

The primary decision rule required the relevant multiplicity-adjusted lower confidence bound to exceed the prespecified practical margin of +0.03. Primary, Sensitivity A, and Sensitivity B formed one three-member family under neutral public labels. Holm-adjusted p-values (Holm, 1979) and rank-specific familywise percentile intervals were recorded for that family; the Primary adjusted interval was {[}0.224173, 0.251047{]}. A positive point estimate alone was therefore insufficient for a \texttt{WIN}. The opportunity-specificity family used the stored \texttt{PAIRED\_FRACTIONAL\_SIGN\_FLIP\_TWO\_SIDED} procedure. We refer to it below as a paired fractional sign-flip test, not an exact enumeration.

The terminal label \texttt{INCONCLUSIVE} means that a contrast did not clear the prespecified practical decision rule. It is not proof of zero effect. Likewise, a confidence interval excluding zero does not by itself establish that the effect exceeded the practical margin after the required adjustment. Statistical citations identify the procedure families; the frozen protocol and evidence records, not those citations, establish the implementation used here.

All families, populations, margins, and terminal decision rules were frozen before the result was finalized. Sensitivity surfaces were interpreted within their own matched comparisons. Cross-sensitivity differences were not converted into a causal configuration trend because the public neutral labels deliberately withhold the underlying axis and the study did not define such a causal contrast as its headline.

\subsection{Frozen analysis and independent verification}\label{frozen-analysis-and-independent-verification}

The Study A result was finalized as an immutable package with separately bound result, retrieval, and graph identities. An independent verifier rehashed the package, confirmed the 7,405-row retrieval/evaluation join, and recomputed the headline paired result, its inferential surfaces, the bridge/comparison strata, the topology strata, and the retained anomaly from the sealed evaluation data. The package did not include the raw source corpus, so the verification did not rerun retrieval from source documents.

\section{Results}\label{results}

\subsection{Primary full-population result}\label{primary-full-population-result}

The Primary relational allocation recovered complete supporting evidence substantially more often than its matched flat baseline. Across 7,405 paired questions, the risk difference was 0.237677 (95\% paired question-level bootstrap interval 0.226874 to 0.248886). The Holm-family adjusted interval was 0.224173 to 0.251047, leaving its lower bound well above the prespecified +0.03 practical margin. The terminal disposition was \texttt{WIN}.

The Primary allocation recovered the complete support set on 4,315 questions (58.3\%), compared with 2,555 questions (34.5\%) for the matched flat baseline. These absolute rates describe each arm; the paired 23.8-point risk difference remains the estimand because both outcomes were observed on the same questions.

The exact decimal serves the audit. In ordinary terms, the observed difference amounts to approximately 24 more complete recoveries per 100 questions. This is a difference in access to the complete official support set, not a 24-point increase in answer accuracy.

The net count provides an independent arithmetic read of the point estimate: 1,952 improvements minus 192 harms yields 1,760 net complete recoveries, and 1,760 divided by 7,405 is 0.237677. This equality does not replace the bootstrap interval or adjusted decision rule. It makes the paired effect tangible and gives a skeptical reader a direct check on the reported magnitude.

The paired counts show how the average was composed. The Primary allocation improved 1,952 questions, tied the baseline on 5,261, and harmed 192. Most questions were ties. Among the discordant cases, improvements substantially outnumbered harms. The 192 harm cases nevertheless rule out any claim that relational allocation is uniformly or monotonically beneficial.

\begin{figure}
\centering
\includegraphics[width=0.96\linewidth,height=\textheight,keepaspectratio,alt={Primary complete-support recovery contrast. The first-read message is that a matched relational allocation produced a large net gain, while 192 questions still moved in the wrong direction. The figure reports n=7,405, absolute recovery rates 58.3\% and 34.5\%, paired risk difference 0.2377, 95\% interval {[}0.2269, 0.2489{]}, the +0.03 practical threshold, and improve/tie/harm counts 1,952/5,261/192.}]{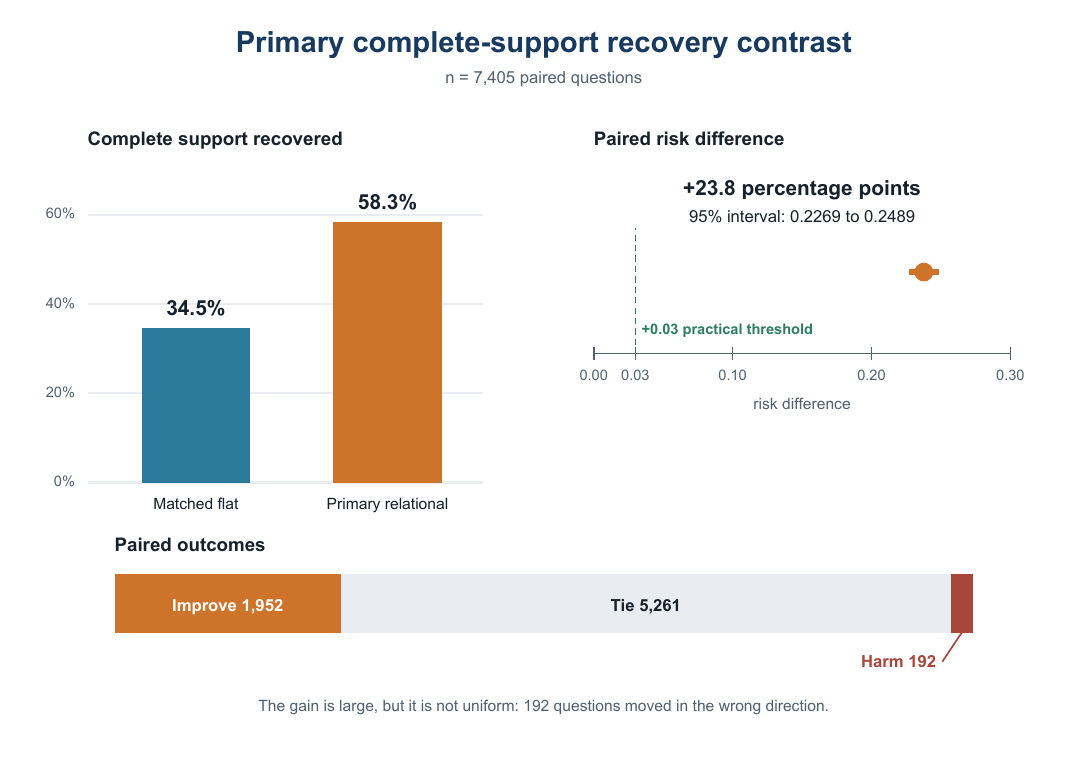}
\caption{Primary complete-support recovery contrast. The first-read message is that a matched relational allocation produced a large net gain, while 192 questions still moved in the wrong direction. The figure reports \protect\texttt{n=7,405}, absolute recovery rates 58.3\% and 34.5\%, paired risk difference 0.2377, 95\% interval {[}0.2269, 0.2489{]}, the +0.03 practical threshold, and improve/tie/harm counts 1,952/5,261/192.}\label{fig:figure2}
\end{figure}

\subsection{Prespecified bridge and comparison strata}\label{prespecified-bridge-and-comparison-strata}

The aggregate effect was strongly heterogeneous across the official question types. Among 5,918 bridge questions, the paired risk difference was 0.286921 (95\% interval 0.274417 to 0.299594), with 1,789 improvements, 4,038 ties, and 91 harms. The adjusted terminal disposition was \texttt{WIN}.

Among 1,487 comparison questions, the paired risk difference was 0.041695 (95\% interval 0.020175 to 0.062542), with 163 improvements, 1,223 ties, and 101 harms. Although the point estimate was positive, its adjusted lower bound was 0.018157, below the prespecified +0.03 practical margin. The correct disposition is therefore \texttt{INCONCLUSIVE}, not a smaller \texttt{WIN}.

The two strata must be read together. Bridge questions constituted the larger stratum and carried most of the aggregate advantage. Comparison questions did not show the same practical separation. The result is not a uniform 23.8-point law across question forms; it is an average dominated by a prespecified regime in which support is often relationally distributed.

The prespecified neighborhood-growth strata were sharply imbalanced. The contained stratum included 7,395 questions and closely tracked the full result: RD 0.237593 (95\% interval 0.226640 to 0.248411), with 1,949 improvements, 5,254 ties, and 192 harms; its adjusted disposition was \texttt{WIN}. The expanding stratum contained only 10 questions: RD 0.300000 (95\% interval 0.000000 to 0.600000), with 3 improvements, 7 ties, and no harms; its adjusted disposition was \texttt{INCONCLUSIVE}. The expanding estimate is too sparse to support a broad topology claim.

Prespecified contained-regime interactions pointed in the same heterogeneous direction without becoming new arm effects. The bridge-contained regime included 2,363 questions and had an interaction risk difference of 0.148774 (Holm-adjusted \texttt{p=8.51×10\^{}-35}). The comparison-contained regime included 1,487 questions and had an interaction risk difference of -0.245227 (Holm-adjusted \texttt{p=3.57×10\^{}-68}). The hub-oriented regime contained no qualifying questions and was classified as empty or underpowered; no effect claim is available from it. These aggregate topology surfaces show effect modification under the frozen features. They do not supply a deployable routing rule or change the bridge and comparison arm estimates reported above.

\begin{figure}
\centering
\includegraphics[width=0.96\linewidth,height=\textheight,keepaspectratio,alt={Prespecified question-type strata. The relational advantage is concentrated in bridge questions. Bridge and comparison estimates appear on the same probability-difference axis with their intervals and adjusted dispositions; the comparison contrast is labeled INCONCLUSIVE under the practical rule.}]{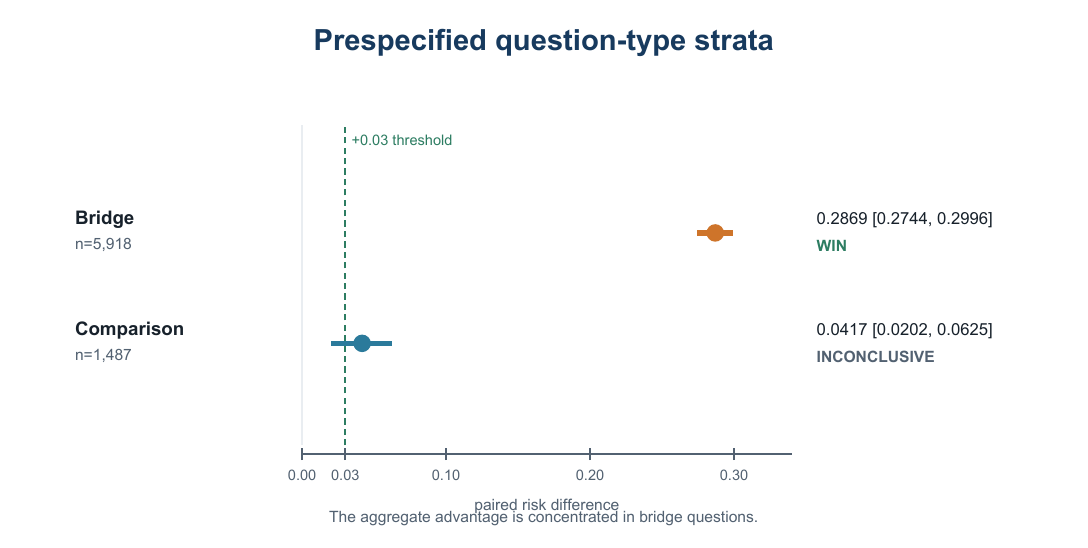}
\caption{Prespecified question-type strata. The relational advantage is concentrated in bridge questions. Bridge and comparison estimates appear on the same probability-difference axis with their intervals and adjusted dispositions; the comparison contrast is labeled \protect\texttt{INCONCLUSIVE} under the practical rule.}\label{fig:figure3}
\end{figure}

\subsection{Relationship-control surfaces}\label{relationship-control-surfaces}

The frozen analysis first reports the graph controls over all 7,405 questions, without oracle conditioning. The real relationship allocation exceeded the random-neighbor control by RD 0.165132 (95\% interval 0.156948 to 0.173261), with 2,017 improvements, 4,595 ties, and 793 harms. It exceeded the degree-preserving shuffled-graph control by RD 0.308602 (95\% interval 0.298231 to 0.319163), with 2,286 improvements, 5,118 ties, and 1 harm. These mandatory full-population surfaces are descriptive: the sealed result does not assign them the adjusted confirmatory disposition used for the separate opportunity-specificity family.

The gold-derived, evaluation-only opportunity population then asks a narrower, conditional question. It contains 3,562 questions selected using official support and reachability in the real graph. Within that restricted diagnostic, the real graph exceeded the random-neighbor control by RD 0.342532. Its multiplicity-adjusted 95\% interval was 0.325632 to 0.359349; the paired fractional sign-flip p-value was 0.00009999 before Holm adjustment and 0.00019998 after adjustment.

The real graph exceeded the degree-preserving shuffled-graph control by RD 0.637339. Its adjusted 95\% interval was 0.621617 to 0.652723, with the same reported pre-adjustment and Holm-adjusted p-values. Both contrasts had a \texttt{WIN} disposition within the prespecified opportunity-specificity family.

Across the full population, randomizing or shuffling relationships did not reproduce the real-graph outcome. The opportunity diagnostic strengthens that separation where the real graph is known after the fact to contain a usable route to gold support. That conditioning is part of the result, not a footnote: it can sharpen the observed contrasts and cannot be used as an operational selector. The diagnostic supports conditional specificity under the frozen design; it does not prove that every full-population gain had the same causal pathway.

The Primary relational-minus-flat contrast inside the opportunity population was itself 0.543515 (95\% interval 0.526109 to 0.560079). We report that value to identify the diagnostic surface, not to advertise it as expected deployment performance. The population was constructed precisely where gold support and graph reachability established an opportunity; its effect is therefore expected to differ from the unconditional 0.237677 full-population estimate. Treating the larger number as the headline would substitute an oracle-assisted population for the actual operating population.

\begin{figure}
\centering
\includegraphics[width=0.96\linewidth,height=\textheight,keepaspectratio,alt={Full-population controls and conditional relationship diagnostic. Panel A reports real-versus-random and real-versus-shuffled differences across all 7,405 questions. Panel B reports the prespecified gold-derived, evaluation-only conditional diagnostic (n=3,562). Panel B is not a deployment estimate and does not replace Figure 2.}]{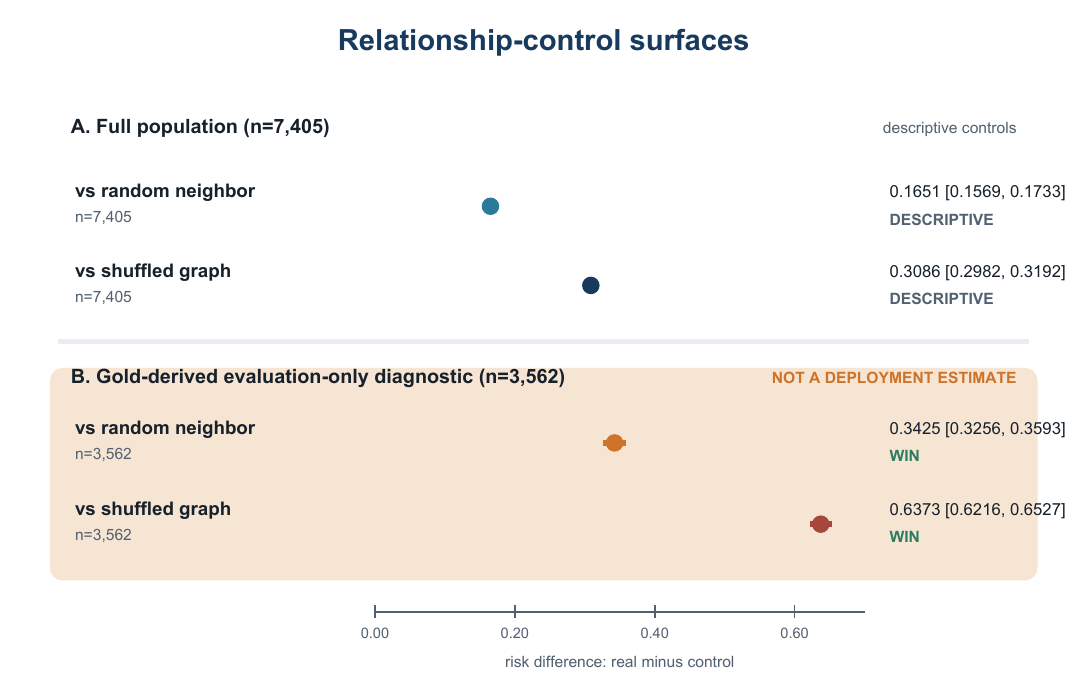}
\caption{Full-population controls and conditional relationship diagnostic. Panel A reports real-versus-random and real-versus-shuffled differences across all 7,405 questions. Panel B reports the prespecified gold-derived, evaluation-only conditional diagnostic (\protect\texttt{n=3,562}). Panel B is not a deployment estimate and does not replace Figure 2.}\label{fig:figure4}
\end{figure}

\subsection{Sensitivity surfaces}\label{sensitivity-surfaces}

The aggregate result remained positive in two neutral-labeled sensitivity comparisons, each evaluated against its own matched baseline. Sensitivity A had a paired risk difference of 0.221202 (95\% interval 0.209858 to 0.232681), with 1,930 improvements, 5,183 ties, and 292 harms. Sensitivity B had a paired risk difference of 0.205537 (95\% interval 0.194058 to 0.217016), with 1,871 improvements, 5,185 ties, and 349 harms. Both received a \texttt{WIN} disposition.

{\def\LTcaptype{none} 
\begin{longtable}[]{@{}
  >{\raggedright\arraybackslash}p{(\linewidth - 10\tabcolsep) * \real{0.1364}}
  >{\raggedleft\arraybackslash}p{(\linewidth - 10\tabcolsep) * \real{0.1818}}
  >{\raggedleft\arraybackslash}p{(\linewidth - 10\tabcolsep) * \real{0.1818}}
  >{\raggedleft\arraybackslash}p{(\linewidth - 10\tabcolsep) * \real{0.1818}}
  >{\raggedleft\arraybackslash}p{(\linewidth - 10\tabcolsep) * \real{0.1818}}
  >{\raggedright\arraybackslash}p{(\linewidth - 10\tabcolsep) * \real{0.1364}}@{}}
\toprule\noalign{}
\begin{minipage}[b]{\linewidth}\raggedright
Public allocation
\end{minipage} & \begin{minipage}[b]{\linewidth}\raggedleft
n
\end{minipage} & \begin{minipage}[b]{\linewidth}\raggedleft
Paired risk difference
\end{minipage} & \begin{minipage}[b]{\linewidth}\raggedleft
95\% interval
\end{minipage} & \begin{minipage}[b]{\linewidth}\raggedleft
Improve / tie / harm
\end{minipage} & \begin{minipage}[b]{\linewidth}\raggedright
Disposition
\end{minipage} \\
\midrule\noalign{}
\endhead
\bottomrule\noalign{}
\endlastfoot
Primary & 7,405 & 0.237677 & {[}0.226874, 0.248886{]} & 1,952 / 5,261 / 192 & \texttt{WIN} \\
Sensitivity A & 7,405 & 0.221202 & {[}0.209858, 0.232681{]} & 1,930 / 5,183 / 292 & \texttt{WIN} \\
Sensitivity B & 7,405 & 0.205537 & {[}0.194058, 0.217016{]} & 1,871 / 5,185 / 349 & \texttt{WIN} \\
\end{longtable}
}

The table shows that a positive aggregate separation was not exclusive to the Primary surface. It does not identify what configuration changed among the neutral labels, establish an ordered trend, or support a causal claim about any held allocation parameter. The labels and table order carry no public mapping to held internal configurations.

The rising harm counts across the three public rows are descriptive facts, not an interpretable public configuration curve. Because the neutral labels disclose no configuration mapping, readers cannot infer which operational choice produced the difference, and the manuscript does not invite them to reverse- engineer it through ordering or visual encoding. What the cleared surfaces do show is narrower: each neutral-labeled allocation retained a positive aggregate matched contrast, while none eliminated question-level harm.

\subsection{Adverse support-resolution anomaly}\label{adverse-support-resolution-anomaly}

One adverse benchmark example is preserved in the result rather than silently discarded. For the Jimmy Butler (basketball) question, one official support reference fell outside the resolved article range. The Primary allocation, its matched baseline, and the evaluation-only oracle diagnostic all failed complete support recovery. The case was therefore non-rescuing: it did not create an artificial Primary improvement, and it remained in the aggregate population. The anomaly is considered further in Section 6 without inferring general benchmark corruption.

\section{Discussion}\label{discussion}

\subsection{What the result establishes}\label{what-the-result-establishes}

Retrieval systems do not merely rank evidence. They turn a ranking into a bounded context, and that conversion is itself a consequential design choice. Under the frozen Study A design, spending a matched allowance on relevance- selected seeds plus bounded graph-adjacent context produced substantially more complete support sets than spending it on additional flat-ranked items. The 23.8-point paired difference is large enough that context allocation should not be treated as implementation detail in this setting.

The controlled comparison is what gives the result its force. A broad systems contest changes models, indexes, training, budgets, rerankers, graph construction, and readers at once. Study A instead holds the common ranking stage and permitted context ceilings fixed and changes the allocation strategy. That narrower design cannot identify the best production system, but it can show that the allocation decision deserves independent measurement.

Complete support recovery is also a meaningful endpoint in its own right. A downstream model cannot use evidence that retrieval never provides. This does not mean that providing the official support guarantees a correct answer. It means that evidence availability is one separable failure surface in a larger retrieval-and-reasoning pipeline. Study A measures that surface directly.

The result rules out one assumption and leaves another untouched. Post-ranking allocation was not innocuous under this design. That is enough to reject its treatment as implementation detail. It is not enough to install the tested allocation as a universal default; that would require broader comparators, corpora, costs, and selection evidence.

\subsection{Where the advantage concentrates}\label{where-the-advantage-concentrates}

The bridge/comparison split sharpens the interpretation. The aggregate advantage was not evenly distributed; it was much larger for bridge questions and practically inconclusive for comparison questions. That pattern is consistent with a simple conjecture: relational expansion may be most useful when an initially relevant item exposes a connection to later support that is not independently prominent in the flat ranking.

The conjecture is plausible, not proven. Study A did not directly measure lexical obscurity as a causal mediator. The benchmark question types are coarse labels, and the interaction between ranking position, graph reachability, support annotation, and allocation can produce the observed heterogeneity in more than one way. The evaluation-only opportunity population uses gold information and cannot solve the operational problem of knowing, before retrieval, when relational allocation should be used.

That unresolved question matters because harm was not evenly avoidable. The Primary allocation displaced context that the baseline needed on 192 questions, and the comparison stratum contained nearly as many harms as improvements. A future system should not merely expand relationally; it should learn or infer when the expected value of expansion exceeds the value of the next flat-ranked item under the same ceiling.

\subsection{What the graph controls add}\label{what-the-graph-controls-add}

Across the full population, the real relationship allocation exceeded both the random-neighbor and degree-preserving shuffled controls. If any neighbor-shaped expenditure were sufficient, the random control would have closed more of the gap. If coarse degree structure without the recorded relationships were sufficient, the shuffled control would have done so. Neither reproduced the real graph's result under the frozen control design.

The gold-derived opportunity diagnostic sharpens that separation where the real graph is known after the fact to contain a route to official support. This is stronger than an untested story about why the headline occurred, but weaker than a full causal decomposition. Conditioning on real-graph reachability can itself enlarge the contrast. The diagnostic does not estimate how often a deployable system can recognize an opportunity, how much of the full-population effect the relationship semantics explain, or how the result transfers to a different graph.

\subsection{Design implications, not universal prescriptions}\label{design-implications-not-universal-prescriptions}

Retrieval evaluations should make context allocation explicit. Reporting a retriever name and a context size can hide materially different ways of spending that context. Two systems can share the same nominal ceiling yet expose downstream readers to different complete evidence sets because one continues down a global ranking while another spends part of the allowance locally around relevant material.

The result does not prescribe a universal graph method. Learned path retrieval, multi-hop dense retrieval, GraphRAG, agentic search, and web retrieval make different tradeoffs and operate over different representations. Some may already internalize the allocation problem in a learned policy. Others may benefit from explicit hybrids. Study A supplies a controlled reason to measure the choice, not a verdict on systems it did not test.

\subsection{Next experiments}\label{next-experiments}

The first next experiment is downstream. The same paired contexts should be given to isolated reader or generator sessions under matched prompts and model conditions. Outcomes should include answer accuracy, grounded citation completeness, unsupported-answer rate, and false-answer behavior. That design would test whether improved evidence availability survives the next stage of the pipeline rather than assuming it does.

The second next experiment is comparative. The bounded relational allocation should be evaluated alongside learned dense, learned path, GraphRAG, agentic, and web-retrieval systems under declared context, latency, construction, and lifecycle budgets. Such a study would answer a different question from Study A: not whether allocation matters against a matched flat control, but where this allocation sits among modern end-to-end alternatives.

Such comparisons should preserve the distinction Study A isolates. A system may outperform another because it ranks candidates better, allocates context better, uses a stronger reader, or spends more resources. Reporting only final answer accuracy would collapse these causes. A useful follow-on should expose candidate ranking, context composition, complete support recovery, answer behavior, latency, and construction cost as separate but connected surfaces.

The third next experiment is selective allocation. A deployable policy must predict opportunity without official support labels. It should be judged not only by additional recoveries but by the context it displaces, the harms it introduces, and whether its selection cost is justified. The present heterogeneity and diagnostic surfaces identify the problem; they do not yet solve it.

\section{Limitations, Reproducibility, and Disclosure}\label{limitations-reproducibility-and-disclosure}

\subsection{Outcome limitation}\label{outcome-limitation}

Study A ends at retrieval. It reports whether the complete official support set was recovered, not whether a reader answered correctly, cited the evidence, learned a fact, revised a belief, or resisted a false revision. It also does not test discovery of information absent from the resident corpus. Any downstream benefit remains a hypothesis for a separately designed experiment.

\subsection{Comparator and generalization limitation}\label{comparator-and-generalization-limitation}

The matched flat baseline was selected to isolate allocation after a common ranking stage. It is not a comprehensive representation of contemporary retrieval. The study contains no direct comparison with learned path retrieval, multi-hop dense retrieval, GraphRAG, agentic search, web retrieval, or a production RAG system. The data come from one frozen HotpotQA FullWiki design and one frozen relationship graph. General superiority is therefore outside the evidence.

HotpotQA itself imposes external-validity limits. Its official support sets are annotations for constructed multi-hop questions over a fixed Wikipedia-derived corpus. Other domains may have denser redundancy, noisier relationships, different document granularity, incomplete graphs, or evidence that cannot be represented as a small official set. The effect may change when evidence is temporal, adversarial, proprietary, or absent from the resident corpus.

\subsection{Heterogeneity and harm}\label{heterogeneity-and-harm}

The average effect is driven primarily by bridge questions. The comparison estimate was positive but did not clear the adjusted practical-margin rule. Across the full population, 192 questions were harmed by the Primary allocation relative to the matched baseline. Sensitivity A and Sensitivity B also retained positive aggregate effects but contained more harm cases than the Primary surface. These findings preclude a monotonic-benefit interpretation and make selective allocation an open research problem.

The prespecified topology interactions reinforce heterogeneity but do not solve selection. One hub-oriented regime was empty, so the study provides no estimate for that condition. The expanding neighborhood-growth stratum contained only 10 questions and was inconclusive despite its positive point estimate. The non- empty interactions are aggregate effect-modification surfaces and cannot be converted into per-query routing decisions without a separate prospective evaluation.

\subsection{Reproducibility layers}\label{reproducibility-layers}

Three reproducibility claims must be distinguished. First, sealed-result recomputation was performed: an independent verifier recomputed the reported results and inferential surfaces from the immutable evaluation package. Second, the final package did not include the raw source corpus needed for an independent source-body rerun, and public materials do not redistribute the corpus or index. Third, the public description omits deterministic allocation details and reconstruction-capable fixtures.

Certain deterministic post-filing allocation details are omitted; the public method is not a full executable replication specification.

The independent recomputation is meaningful evidence of result identity and numerical consistency. It is not equivalent to an unconstrained public replication from source documents. The limitation is substantive and should be considered when evaluating the result.

Public transparency is also asymmetrical. The paper can expose exact reported effects, uncertainty, arm-neutral sensitivities, aggregate graph identity, and the verification route while withholding allocation mechanics that would make the procedure executable. Readers can audit whether the public claims match the sealed result through authorized verification records, but they cannot fully stress-test implementation choices from the public description alone. The manuscript must not call that condition ``fully reproducible.''

\subsection{Lifecycle and graph-construction limits}\label{lifecycle-and-graph-construction-limits}

Budget matching applies to the frozen maximum object and token ceilings. The study does not establish identical realized context use on every question, matched latency, training cost, index cost, graph-construction cost, energy consumption, maintenance burden, or production complexity. The graph identity and aggregate counts establish what result surface was tested; they do not establish that the graph is economical or operationally superior. Because the public description does not resolve which relationship semantics carried the effect, readers cannot evaluate their transfer beyond the frozen corpus.

The inferential surfaces also answer narrower questions than their numerical precision may suggest. Bootstrap intervals quantify uncertainty under the frozen resampling design; they do not capture every uncertainty introduced by benchmark construction, graph choice, resolver behavior, or transfer to another corpus. Holm adjustment controls the prespecified families reported here, not every interpretation a reader might form after seeing the results. The practical margin encodes a frozen decision rule rather than a universal threshold for retrieval importance.

\subsection{Adverse anomaly}\label{adverse-anomaly}

In the Jimmy Butler (basketball) benchmark example, one official support reference fell outside the resolved article range. The Primary allocation, its matched baseline, and the evaluation-only oracle diagnostic all failed complete support recovery. The example remained in the aggregate population and was adverse and non-rescuing; it did not manufacture a win for the relational allocation. One example does not establish general benchmark corruption, but it does demonstrate that support-resolution integrity is part of the measurement surface.

More generally, the binary endpoint inherits the benchmark's support annotations and resolver behavior. An official support set may be sufficient without being unique, and resolver failures can turn an annotation problem into a retrieval failure. The retained anomaly is evidence that this risk is real, not evidence that it is widespread. A future robustness analysis should separate missing evidence, unresolved references, alternate sufficient support, and genuine retrieval misses under a prespecified adjudication protocol.

\subsection{Patent and disclosure relationship}\label{patent-and-disclosure-relationship}

This study evaluates an abstract graph-adjacent context-selection method related to subject matter disclosed in pending U.S. Patent Application No.~19/362,426, ``System M.2: Reflexive Memory Governance and Dual-Bounded Recall Protocols.''

That statement records a disclosure relationship. Patent pendency is not scientific validation and supports no conclusion about patent scope, validity, infringement, licensing, or issued rights. No official public bibliographic record was located during source inspection, so no patent citation is attached.

\section{Conclusion}\label{conclusion}

A retrieval budget can be fixed and still be spent badly. Ranking determines which evidence appears promising; allocation determines which of that evidence, and which material around it, is allowed into the final context. Study A tested that second decision directly.

Under the frozen HotpotQA FullWiki design, allocating the same maximum number of retrieval units and tokens to relevance-selected seeds plus bounded graph-adjacent context increased complete supporting-evidence recovery by 23.8 percentage points over additional flat-ranked context across 7,405 questions. The advantage was concentrated in bridge questions, while the comparison result was practically inconclusive. The real graph exceeded random and shuffled controls over the full population and separated further in a gold-derived, evaluation-only conditional diagnostic, yet 192 full-population questions were harmed.

Those boundaries make the result more specific, not less useful. The study does not show that a downstream model answered better, that every graph expansion is beneficial, or that the tested method outranks modern retrieval systems. It shows that retrieval quality depends on more than which items score highly. It also depends on how a finite context budget is allocated around what the system has already found relevant. The next scientific question is whether that more complete evidentiary substrate produces better grounded answers, and whether a system can learn when relational allocation is worth the context it displaces.

The immediate lesson is therefore neither ``use graphs'' nor ``ranking is obsolete.'' It is to stop treating the step between ranking and context as neutral. Once context is bounded, allocation becomes part of retrieval, and it should be designed, measured, and challenged as such.

A ranking can be strong while the final context remains evidentially incomplete. Conversely, a modest relational step can be consequential when it spends the same finite allowance on evidence the ranking alone would leave beyond the boundary.

\section*{References}\label{references}
\addcontentsline{toc}{section}{References}

Asai, A., Hashimoto, K., Hajishirzi, H., Socher, R., and Xiong, C. (2020). Learning to retrieve reasoning paths over Wikipedia graph for question answering. \emph{International Conference on Learning Representations}. \url{https://openreview.net/forum?id=SJgVHkrYDH}

Edge, D., Trinh, H., Cheng, N., Bradley, J., Chao, A., Mody, A., Truitt, S., Metropolitansky, D., Ness, R. O., and Larson, J. (2024). From local to global: A Graph RAG approach to query-focused summarization. \emph{arXiv:2404.16130}. \url{https://arxiv.org/abs/2404.16130}

Efron, B. (1979). Bootstrap methods: Another look at the jackknife. \emph{The Annals of Statistics, 7}(1), 1--26. \url{https://doi.org/10.1214/aos/1176344552}

Holm, S. (1979). A simple sequentially rejective multiple test procedure. \emph{Scandinavian Journal of Statistics, 6}, 65--70. \url{https://www.jstor.org/stable/4615733}

HotpotQA. (2019). \emph{Official HotpotQA repository} (commit {\footnotesize\texttt{3635853403a8735609ee997664e1528f4480762a}}). \url{https://github.com/hotpotqa/hotpot/tree/3635853403a8735609ee997664e1528f4480762a}

Lewis, P., Perez, E., Piktus, A., Petroni, F., Karpukhin, V., Goyal, N., K\"{u}ttler, H., Lewis, M., Yih, W.-t., Rockt\"{a}schel, T., Riedel, S., and Kiela, D. (2020). Retrieval-augmented generation for knowledge-intensive NLP tasks. \emph{Advances in Neural Information Processing Systems, 33}. \url{https://proceedings.neurips.cc/paper/2020/hash/6b493230-Abstract.html}

McNemar, Q. (1947). Note on the sampling error of the difference between correlated proportions or percentages. \emph{Psychometrika, 12}(2), 153--157. \url{https://doi.org/10.1007/BF02295996}

Robertson, S. E., and Zaragoza, H. (2009). The probabilistic relevance framework: BM25 and beyond. \emph{Foundations and Trends in Information Retrieval, 3}(4), 333--389. \url{https://doi.org/10.1561/1500000019}

Schuirmann, D. J. (1987). A comparison of the two one-sided tests procedure and the power approach for assessing the equivalence of average bioavailability. \emph{Journal of Pharmacokinetics and Biopharmaceutics, 15}(6), 657--680. \url{https://doi.org/10.1007/BF01068419}

Xiong, W., Li, X. L., Iyer, S., Du, J., Lewis, P., Wang, W., Mehdad, Y., Yih, W.-t., Riedel, S., Kiela, D., and O\u{g}uz, B. (2021). Answering complex open-domain questions with multi-hop dense retrieval. \emph{International Conference on Learning Representations}. \url{https://openreview.net/forum?id=EMHoBG0avc1}

Yang, Z., Qi, P., Zhang, S., Bengio, Y., Cohen, W. W., Salakhutdinov, R., and Manning, C. D. (2018). HotpotQA: A dataset for diverse, explainable multi-hop question answering. In \emph{Proceedings of the 2018 Conference on Empirical Methods in Natural Language Processing} (pp.~2369--2380). Association for Computational Linguistics. \url{https://doi.org/10.18653/v1/D18-1259}

\begingroup\small\setlength{\parskip}{0.35em}
\section*{Declarations}\label{declarations}
\addcontentsline{toc}{section}{Declarations}

\subsection*{Data and materials availability}\label{data-and-materials-availability}
\addcontentsline{toc}{subsection}{Data and materials availability}

The official HotpotQA acquisition, format, and license surface is cited above (HotpotQA, 2019). Section 3.4 reports sanitized result, retrieval, and graph identity hashes together with aggregate graph counts. This paper does not distribute raw or processed corpus bytes, private retrieval traces, reconstruction-capable fixtures, or custody coordinates.

\subsection*{Independent verification}\label{independent-verification}
\addcontentsline{toc}{subsection}{Independent verification}

An independent verifier rehashed the immutable package and recomputed the headline paired result, inference, prespecified strata, topology summaries, and retained anomaly from the sealed evaluation data. The source corpus was not included, so this was not a public source-to-result rerun.

\subsection*{AI-assistance disclosure}\label{ai-assistance-disclosure}
\addcontentsline{toc}{subsection}{AI-assistance disclosure}

AI-assisted editorial tools were used for drafting support, source organization, adversarial review, copyediting, and manuscript assembly under the author's direction. Thomson D. Nguy retained the scientific conception, all authorship decisions, and responsibility for the manuscript and its claims.

\subsection*{Competing interests and patent relationship}\label{competing-interests-and-patent-relationship}
\addcontentsline{toc}{subsection}{Competing interests and patent relationship}

The patent relationship is disclosed in Section 6.7. Patent pendency is not scientific validation and supports no conclusion about scope, validity, infringement, licensing, or issued rights.
\endgroup

\end{document}